\documentstyle[amssymb,12pt]{article}
\font\tmsb=msbm10 at12pt
\font\smsb=msbm7
\font\ssmsb=msbm5
\newfam\msbfam
\textfont\msbfam=\tmsb
\scriptfont\msbfam=\smsb
\scriptscriptfont\msbfam=\ssmsb
\def \mth {\fam\msbfam}
\def \Mth#1 {{\mth #1}}

\def \NN {{\Bbb {N}}}
\def \ZZ {{\Bbb {Z}}}
\def \CC {{\Bbb {C}}}

\newcommand \bd {\begin {displaymath}}
\newcommand \ed {\end {displaymath}}

\newcommand \mf {\mathfrak}
\newcommand \pa {\partial}
\newcommand \al {\alpha}
\newcommand \be {\beta}
\newcommand \da {\delta}

\newcommand \ep {\epsilon}
\newcommand \sa {\sigma}

\newcommand \La {\Lambda}
\newcommand \G  {\mathfrak g}
\newcommand \HH {\mathfrak {h}}
\newcommand \SL {{\mathfrak s}{\mathfrak l}}
\newcommand \cF {{\cal F}}
\newtheorem {prop} {Proposition} [section]
\newtheorem {defi} {Definition} [section]
\newtheorem {lem} {Lemma} [section]
\newtheorem {th} {Theorem} [section]

\newtheorem {rk} {Remark}
\newtheorem {nota} {Notation}

\title{The Quantization of the Generalized mKdV Equations 
 for $\widehat {\SL}_2$}
\author{A.  Balan\\
\'Ecole Polytechnique\\
Centre de Math\'ematiques\\
UMR 7640 du CNRS\\
F-91128 PALAISEAU Cedex\\
{\it email}: balan@math.polytechnique.fr}
\date{}
\begin{document}
\maketitle

\abstract{We construct quantum deformations of the 
 integrals of motion of the generalized mKdV equations
 for $\widehat {\SL}_2$. For this,
 we give the relevant vertex operator algebra
 and prove quantum Serre relations for vertex operators, it allows
 to construct a $q$-BGG resolution and to deform the classical
 integrals of motion in a commutativ family.}
 
\newpage

\section{Introduction}

The generalized mKdV equations were introduced in \cite{GS}. They
 are associated to an arbitrary Kac-Moody algebra and
 an integer $\ell$. The Lax operator has the form:
\begin{equation}
{\cal L} = \pa_x + p_{-\ell} +
\sum_{-\ell \leq i \leq 0} \sum_p u_{i,p}(x) e_{i,p} = \pa_x+L,
\end{equation}
 where for all $i$, $(e_{i,p})_p$ is a basis of the principal degree
 $i$ part of the affine Lie algebra
 ${\mf g}$.  In \cite{B}, we studied the simplest case of these equations -- 
when ${\mf g}= \widehat {{\mf s}{\mf l}}_2$ and $\ell = 3$. In that case,
 $L =p_{-3}+ H_{-2}h_{-2}+ E_{-1}e_{-1}+F_{-1}f_{-1}+H_0 h_0$
(we use the convention that $x_n$ has principal degree $-n$).
 The main theorem was a geometric proof of 
the commutation of the classical integrals of motion (already proved in
 \cite{DFG}). For that aim, one of our
 results was that the group-theoretic interpretation of \cite{FF} \cite{FF3}
 -- interpretation of the space of jets of $H_0,E_{-1},F_{-1},H_{-2}$
 with the functions on a
 quotient $N_+/A_+$ of subgroups of the Lie group associated with
 $\widehat {{\mf s}{\mf l}}_2$ -- could be carried out in this situation,
 if we set the 
 central element $d_{-1} = (E_{-1}+F_{-1}+H_{-2}^2)/2$ and all its derivatives
 equal to zero. The action of the derivation $\pa_x$ on:
\bd
\pi_0^0= \CC[H_0^{(n)},E_{-1}^{(n)},F_{-1}^{(n)},H_{-2}^{(n)}]/(d_{-1},d'_{-1},
 \ldots, d^{(n)}_{-1}, \ldots),
\ed
 was then identified to the right action of $p_{-3}$,
 the principal degree $-3$ element of $\widehat {\SL}_2$, on $N_+/A_+$,
 where $N_+$ is viewed as a submanifold of the flag space of $\widehat {\SL}_2$.
 This allowed us, by studying a BGG complex
 (defined in \cite{BGG} \cite{BGG2}), to give a cohomological 
 interpretation of the integrals of motion similar to that of \cite{FF}. 
 We have then showed in \cite{B} the commutation of the classical integrals of
 motion. 
{\th .
 The $\oint d_{2i+1}$, which are in $\CC [ \pi_0^0]$,
 are in involution for the Poisson brackets.}

The $d_{2i+1}$ are some polynomials in $E_{-1},F_{-1},H_{-2},H_0$ of
 the generalised mKdV equations, the $\oint d_{2i+1}$
 form the classical integrals of motion,
 which commute with each other as in \cite{DFG},
 and so generate a family of commuting vector fields.

 In the present paper, we study the quantization of the system and 
we construct quantum deformations of the classical 
 integrals of motion. The main theorem is (theorem \ref{intmot}):

{\th .  
The integrals of motion $\oint d_{2i+1}$ admit quantum deformations, which
 belong to a completed universal envelopping algebra and which
 commute with each other and have the same degrees as the integrals
 of motion.} 

 For this, we construct a Lie algebra 
 naturally associated to the classical Poisson algebra and show,
 only after taking its quotient by the quantum deformation of
 $d_{-1}$ (definition \ref{d-1q}) 
and derivatives, a VOA structure (theorem \ref{VOA}), the three
 axioms for a VOA (defined in \cite{FF}) are proved, field-state correspondence
 (proposition \ref{fs}), Sugawara field (proposition \ref{der}),
 locality (proposition \ref{loc}).
The classical screening charges are replaced by vertex operators
 (introduced in \cite{Bo} \cite{FLM}) 
 $\bar V_+ =  \sum_{i \in \ZZ} E_{-1}[-2/3+i] \otimes V_{+}[-1-i] $
 and $\bar V_- =  \sum_{i \in \ZZ} F_{-1}[-2/3+i] \otimes V_{-}[-1-i]$.
 We show (theorem \ref{qSerre})
 that $\bar V_+$ and $\bar V_-$ satisfy the quantum Serre relations.
 This generalizes a result of \cite{BCP}; this result is obtained by a
 reasoning using the analogy with a lattice model. 
 Finally, we construct a quantum BGG resolution and use
 deformation arguments to show that the classical integrals of motion
 admit quantizations, which commute to each other (theorem \ref{intmot}).
 
\section{The Lie algebra $\tilde {\G}$, associated fields}
\subsection{The Lie algebra $\tilde {\G}$}
The Poisson brackets at level of the variables of the generalized
 mKdV equations are the following ones in \cite{B} \cite{DFG}:
\bd
\{ H_0 (x), H_0 (y) \} = {\frac {1}{2}} \pa_{x} \da_{x,y},
\ed
\bd
\{ H_0 (x), E_{-1} (y) \} = 0, 
\{ H_0 (x), F_{-1} (y) \} = 0,
\{ H_0 (x), H_{-2} (y) \} = 0, 
\ed
\bd
\{ E_{-1} (x) , E_{-1} (y) \} =0,
\{ F_{-1}(x) , F_{-1}(y) \} =0,
\ed
\bd
\{ E_{-1}(x), F_{-1}(y) \} = -4 H_{-2}(x) \da_{x,y},
\ed
\bd 
\{ E_{-1}(x), H_{-2}(y) \} = 2 \da_{x,y},
\{ F_{-1}(x), H_{-2}(y) \} = - 2 \da_{x,y},
\ed
\bd
\{ H_{-2}(x), H_{-2}(y) \} = 0.
\ed
The Lie algebra ${\G}$ is defined by its generators,
 $E_{-1}[n]$, $F_{-1}[n]$ and $H_{-2}[n]$:
\bd
 E_{-1}[n], n \in -2/3 + \ZZ, 
\ed
\bd
 F_{-1}[n], n \in -2/3 + \ZZ, 
\ed 
\bd
 H_{-2}[n], n \in -1/3 + \ZZ,
\ed
and relations, with $\be$ a deformation parameter:
\bd
 [ E_{-1}[n], F_{-1}[m] ] = -4 \be H_{-2} [n+m], 
\ed 
\bd
 [ E_{-1}[n], H_{-2}[m] ] = 2 \be \da_{n+m,0},
 [ F_{-1}[n], H_{-2}[m] ] = - 2 \be \da_{n+m,0}, 
\ed 
\bd
 [H_{-2}[n], H_{-2}[m] ] = 0.  
\ed 
Let $\HH$ be the algebra generated by:
\bd
 H_0[n], n \in \ZZ,
\ed
and relations:
\bd
 [H_0[n], H_0[m]]= (1/2)(n) \be \da_{n + m , 0}.
\ed
The algebra $\tilde {\G}$ is the direct product of $\G$ and $\HH$.
\bd
 [ H_0 [n], E_{-1} [m] ] = 0, [ H_0 [n], F_{-1} [m] ] =0,
 [ H_0 [n], H_{-2} [m] ] = 0.
\ed 
The corresponding fields are:
\bd 
E_{-1}(z)= \sum_{n \in -2/3+\ZZ} E_{-1}[n]z^{-n-2/3}, 
\ed 
\bd
F_{-1}(z) =\sum_{n \in -2/3+\ZZ} F_{-1}[n]z^{-n-2/3}, 
\ed 
\bd
H_{-2}(z)= \sum_{n \in -1/3+\ZZ} H_{-2}[n]z^{-n-1/3},
\ed
\bd
H_0(z) = \sum_{n \in -1+\ZZ} H_0[n]z^{-n-1}.
\ed
 The sums are over the ring $\ZZ$ shifted by a 
 certain number such that $z$ has an integer power.

\subsection{The normal ordering of two fields}

The normal ordering of two fields $A,B$ of conformal weights $\Delta_A$ and
 $\Delta_B$, is defined by the following way:

\bd
A(z) = \sum_{n \in \ZZ - \Delta_A} A[n] z^{-n - \Delta_A},
\ed
\bd
B(z) = \sum_{n \in \ZZ - \Delta_B} B[n] z^{-n - \Delta_B},
\ed
\bd
A_{reg}(z)= \sum_{-n - \Delta_A \geq 0} A[n] z^{-n - \Delta_A},
\ed
\bd
A_{sing}(z)= \sum_{-n - \Delta_A < 0} A[n] z^{-n - \Delta_A},
\ed
\bd
:AB:(z) = A_{reg} (z)B(z)+ B(z) A_{sing}(z).
\ed

\section{$d_{-1}^{\be}$ quantum central element}

\subsection{Definition of $d_{-1}^{\be}$ }

The following element is defined:
{\defi . \label{d-1q}
\begin{equation}
d_{-1}^{\be} (z) =  E_{-1}(z) + F_{-1}(z) + : H_{-2} H_{-2} :(z).
\end{equation}}

\subsection{Centrality}

{\lem . The Fourier coefficients of $d_{-1}^{\be}$ are central 
 in the enveloping algebra $U\G$.}

Indeed, the normal ordering is developped and gives 
a decomposition of the delta function.

{\defi . Let $U {\G}^{(0)}$ be the quotient
 of $U\G$ by the ideal generated by the $d_{-1}^{\be}[n]$.}

\medskip

\subsection{The degree of $d_{-1}^{\be}$}

The element $d_{-1}^{\be}$ is homogeneous:
\bd
 d_{-1}^{\be} = E_{-1} +F_{-1} + : H_{-2}H_{-2}:, 
\ed
 as, $deg(E_{-1})=deg(F_{-1}) = -2,deg(H_{-2})=-1$, it gives: 
\bd
deg(d_{-1}^{\be}) = -2.  
\ed
The conformal weight is $2/3$.

\section{The VOA  associated with ${\G}$}

The goal of this section is to show the following theorem:

{\th . There exists a VOA structure associated with $\G$
 and $d_{-1}^{\be}$. \label{VOA}}

The vacuum module is $\pi_0^{\be}$ and the space of fields $\cal F$,
 as defined above \cite{FF} \cite{FLM}.

\subsection{The quantum module of the states}

{\defi .  Let $U^+$ be the sub-algebra of $(U{\G})^{(0)}$
 generated by the $X[n], n \geq 0, X = E_{-1}, F_{-1}, H_{-2}$
 and  let $U^-$ be the one generated by the $X[n],n<0, X = E_{-1},F_{-1},
 H_{-2}$.}

The quantum module $\pi_0^{\be}$ is defined by:

{\defi .
\bd
\pi_0^{\be}= (U{\G})^{(0)} \otimes_{U_+} \CC \cong {U^-},
\ed
with $U{\G}_+$ acting trivialy
 over $\CC$ as a character, putting all $X[n]$ to zero.}

Let $| \Omega \rangle$ be the following element:
\bd
| \Omega \rangle = 1\otimes 1,
\ed
then the application  $U^- \rightarrow \pi_0^{\be}$,
 $T \mapsto T | \Omega \rangle$ is an isomorphism of vector spaces.

A degree is defined over $\pi_0^{\be}$ by:

\bd
deg(|\Omega \rangle )=0, deg(E_{-1}[n])=n, deg(F_{-1}[n])=n, deg(H_{-2}[n])=n.
\ed

\subsection{The quantum algebra of the fields}

The algebra of the fields $\cal F$ is defined by:

\bd
{\cal F} \hookrightarrow \widehat {U {\G}}^{(0)}[[z,z^{-1}]],
\ed

where $\cal F$ is freely generated by $E_{-1}(z)$, $F_{-1}(z)$ and $H_{-2}(z)$,
 and the derivatives of the fields, taking the normal orderings in
 the completed space of ${U {\G}}^{(0)}[[z,z^{-1}]]$.
 As $d_{-1}^{\be}$ is homogeneous, a degree is defined over $\cal F$ by:

\bd
deg(E_{-1}(z)) = -2/3, deg(F_{-1}(z))= -2/3, deg(H_{-2}(z))= -1/3,
\ed
\bd
deg(:AB:)= deg(A) + deg(B),
\ed
and:
\bd
deg(A') = deg(A)-1.
\ed

The degrees introduced in the \cite{B} are $-3$ times the one
 defined here. The conformal weight is the opposite of the above degree.

{\prop . The normal product from the left to the right of the following
 expressions:

\bd
{E_{-1}(z)}^{e_0} {E_{-1}'(z)}^{e_1}...{E_{-1}^{(n)}(z)}^{e_n}...
\ed
\bd
{F_{-1}(z)}^{f_0} {F_{-1}'(z)}^{f_1}...{F_{-1}^{(n)}(z)}^{f_n}...
\ed
\bd
{H_{-2}(z)}^{h_0} {H_{-2}'(z)}^{h_1}...{H_{-2}^{(n)}(z)}^{h_n}... ,
\ed
define a basis of $\cal F$.}

{\it Proof}: 
this proposition is proved using associativity and commutativity of 
 normal ordered products, up to terms of lower degree.

\subsection{The fields-states correspondence}

It is first given the following lemma:

{\lem . If both limits $lim_{z \rightarrow 0} A(z)|\Omega\rangle$ and
 $lim_{z \rightarrow 0} B(z)|\Omega\rangle$ exist,
 then the limits $lim_{z \rightarrow 0} :AB:(z)
 |\Omega\rangle$ and $lim_{z \rightarrow 0} A'(z) | \Omega \rangle$ also exist.}

{\prop . There exists a correspondence: \label{fs}
\bd
\pi_0^{\be} \; \; \; \; \; {\stackrel {\phi}{\cong}} \; \; \; \; \; {\cal F},
\ed
\bd
 X \; \; \; \; \; \mapsto \; \; \; \; \; \phi_X,
\ed
so that:
\bd
lim_{z \rightarrow 0} \phi_X(z) | \Omega\rangle= X .
\ed}

A derivation is defined over the fields by the usual derivation
 and too for the states $v$ by:

\bd
\pa | \Omega \rangle =0,
\ed
\bd
\pa(E_{-1}[n]v) = (-n+1/3) E_{-1}[n-1]v + E_{-1}[n] \pa v,
\ed
\bd
\pa(F_{-1}[n]v) = (-n+1/3) F_{-1}[n-1]v + F_{-1}[n] \pa v,
\ed
\bd
\pa(H_{-2}[n]v) = (-n+2/3) H_{-2}[n-1]v + H_{-2}[n] \pa v.
\ed

It gives:

\begin{equation}
\frac {\pa}{\pa z} \circ \phi = \phi \circ \pa.
\end{equation}

\subsection{The derivation of the VOA}

Set:
\begin{equation}
T(z) = -1/4 [:E'_{-1}H_{-2}:(z)- :F'_{-1}H_{-2}:(z)],
\end{equation}
\bd
T(z)= \sum_n L_n z^{-n-2}.
\ed

{\prop .   \label{der}
\bd
[L_0, X(z) ] = X'(z) , X \in {\cal F}.
\ed}
The computation is standard but uses ${d'_{-1}}^{\be}=0$.

{\rk .
The field $T(z)$ obeys the relations of the Virasoro algebra without
 central charge.}

\subsection{The locality of the fields of the VOA}

{\prop . All fields  of the space $\cF$ 
are local with respect to each other. \label{loc}}

The two fields $E_{-1}$ and $F_{-1}$ are considered,
 and their correlation function over two vectors $| \xi \rangle$
 and $|v \rangle$.
\bd
\langle \xi|:E_{-1}(z)F_{-1}(w): |v \rangle =
\ed
\bd
\langle \xi |{E_{-1}}_{reg}(z)F_{-1}(w) +
 F_{-1}(w) {E_{-1}}_{sing}(z) |v \rangle,
\ed
\bd
\langle \xi | E_{-1}(z) F_{-1}(w)| v \rangle = 
\langle \xi | :E_{-1}(z) F_{-1}(w):| v \rangle +
\ed
\bd
 \sum_{-n -2/3< 0 ,m} \langle \xi |[E_{-1}[n],F_{-1}[m]]| v
 \rangle  z^{-n-2/3}w^{-m-2/3},
\ed
with the formula of the  commutator,
\bd
\sum_{-n-2/3<0,m}
 \langle \xi |(-4 \be H_{-2}[n+m])| v \rangle  z^{-n-2/3}w^{-m-2/3},
\ed
the sum is the expansion for $w$ small with respect to $z$ of:
\bd
\langle \xi|H_{-2}(w)| v \rangle /(z-w),
\ed
then the same holds with:
\bd
\langle \xi |F_{-1}(w)E_{-1}(z)| v \rangle,
\ed
for $z$ chosen small with respect to $w$.
Therefore, both expressions $\langle \xi |E_{-1}(z)F_{-1}(w)| v
 \rangle $ and $\langle \xi |F_{-1}(w) E_{-1}(z)|v \rangle $
 give expansions of the same rational function
 on $\CC^2$ in different spaces of formal series.
 This proves the axiom of locality.

\subsection{The VOA associated with $\tilde {\mf g}$}

{\defi . Let $\HH$ be the algebra generated by the $H_0[n]$.
 Let $\pi_0^{\HH ,\beta}$ be the Fock vacuum  module.}

The tensor product:
\bd
\tilde \pi^{\beta}_0 = \pi_0^{\HH, \beta} \otimes \pi^{\beta}_0,
\ed
admits a VOA structure.

It exists a VOA associated with $\tilde {\mf g}$ 
which is the tensor product of the VOA associated with $\mf g$ by the
 one attached to $\HH$, it possesses a vacuum module
 $\tilde \pi_0^{\be}$ and a space of fields $\tilde {\cal F}$.

The Virasoro field of the algebra tensor product is:
\bd
T_0(z) \otimes 1 + 1\otimes T(z),
\ed
with $T_0(z)$, the Virasoro field of the free fields \cite{FF}.

\section{The quantum  Serre relation}

\subsection{The vertex operators}

 We will define quantum analogues $\bar V_\pm$ of the classical
 screening charges $Q_+$ and $Q_-$ which appeared in \cite{B}.

{\defi .
Let the algebra ${\mf g}'$ be the semidirect product of
 ${\mf g}$ by $p$ so that:
\bd
[H_0[0],p]=1,
\ed
\bd
[X[n],p]=0,
\ed
for $X[n] \neq H_0[0]$. Let the algebra ${{\mf g}'}_+$ be
 constructed by $H_0[0],p,X[n]$, for positiv $n$.}

{\defi .
 A character $\chi_n$ is defined over this algebra by:
\bd
p \mapsto n
\ed
\bd
H_{0}[0] \mapsto 0,
\ed
\bd
X[n] \mapsto 0.
\ed}

{\defi .
The quantum modules $\tilde \pi_n^{\beta}$ are:
\bd
\tilde \pi_n^{\beta}=
 U{\tilde { \mf g}'} \otimes_{U{\tilde {{\mf g}'}_+}} \CC_{\chi_n}.
\ed}

{\defi .
Two operators $\bar V_+$ and $\bar V_-$ are defined as endomorphisms
 of each module of highest weight for $\tilde {\mf G}$, the
 semi-direct product of the envelopping algebra of $\tilde {\G}$
 by $e^{\pm p}$, elements which commute with all the generators
 of $\tilde {\G}$ excepted $H_0[0]$, and such that:
\bd
 e^{\pm p}H_0[0]e^{\mp p} = H_0[0] \pm  \beta.
\ed
The sum of $\tilde \pi^{\beta}_{n}$ is such a module of highest weight.} 

\bd
h_0 (z)= \int^z H_0,
\ed
\bd
V_+(z)=  e^{h_0^+(z)}e^p e^{h_0^-(z)},
\ed
with $h_0^\pm$:
\bd
h_0^+(z) = - \sum_{n \in - {\NN}^*} {H_{0}[n] \over n} z^{-n},
\ed
\bd
h_0^-(z) = - \sum_{n \in  {\NN}^*} {H_{0}[n] \over n} z^{-n},
\ed
\bd
\bar V_+=\int E_{-1}(z) V_+(z) dz=
\ed
\bd
 =\int \sum_n E_{-1}[n]z^{-n-2/3} V_+(z)dz,
\ed
with:
\bd
V_+(z)= \sum_n V_+[n]z^{-n},
\ed
\bd
\bar V_-=\int E_{-1}(z) V_-(z) dz=
\ed
\bd
 =\int \sum_n E_{-1}[n]z^{-n-2/3}V_-(z)dz,
\ed
with:
\bd
V_-(z)= \sum_n V_-[n]z^{-n},
\ed
\bd
\bar V_{+} = \sum_{i \in \ZZ} E_{-1}[-2/3+i] \otimes V_{+}[-1-i],
\ed
\bd
\bar V_{-} = \sum_{i \in \ZZ} F_{-1}[-2/3+i] \otimes V_{-}[-1-i].
\ed

\subsection{The quantum Serre relation}

The following quantum Serre relation ($q$-Serre) \cite{CP} between
the operators $\bar V_+$ and $\bar V_-$ is:

{\th . The following quantum Serre relation holds: \label{qSerre}
\begin{equation}
 S= \bar V_{-}\bar V_{+} ^3 - [3] \bar V_{+} \bar V_{-} \bar V_{+}^2 + [3] \bar
V_{+}^2 \bar V_{-} \bar V_{+} - \bar V_{+}^3 \bar V_{-}=0, 
\end{equation}}

with:
\bd 
[3] = q + q^{-1} +1, 
\ed

{\it Proof}:
it gives four terms:
\bd
 S= S_1 -[3] S_2 + [3] S_3 - S_4, 
\ed 
\bd
 S_1 = \sum_{i,j_1,j_2,j_3 \in \ZZ}
E_{-1}[i-2/3]F_{-1}[j_1-2/3]F_{-1}[j_2-2/3]F_{-1}[j_3-2/3] \otimes
\ed
\bd
\otimes V_{-} [-i-1]
V_{+}[-j_1-1]V_{+}[-j_2-1]V_{+}[-j_3-1], 
\ed 
\bd 
S_2 = \sum_{i,j_1,j_2,j_3 \in \ZZ}
 F_{-1}[j_1-2/3]E_{-1}[i-2/3]F_{-1}[j_2-2/3]F_{-1}[j_3-2/3] \otimes
\ed
\bd
\otimes
V_{+}[-j_1-1] V_{-}[-i-1]V_{+}[-j_2-1]V_{+}[-j_3-1],
\ed 
\bd 
S_3 =
\sum_{i,j_1,j_2,j_3 \in \ZZ}
F_{-1}[j_1-2/3]F_{-1}[j_2-2/3]E_{-1}[i-2/3]F_{-1}[j_3-2/3] \otimes
\ed
\bd
\otimes
 V_{+}[-j_1-1]V_{+}[-j_2-1]V_{-}[-i-1]V_{+}[-j_3-1], 
\ed 
\bd 
S_4 = \sum_{i,j_1,j_2,j_3 \in
  \ZZ} F_{-1}[j_1-2/3]F_{-1}[j_2-2/3]F_{-1}[j_3-2/3]E_{-1}[i-2/3] \otimes
\ed
\bd
\otimes
V_{+}[-j_1-1]V_{+}[-j_2-1]V_{+}[-j_3-1]V_{-} [-i-1].  
\ed
It is then possible to reorganize the four terms using the relations
 of the Lie algebra.

A family of elements of $(U \tilde {\G})^{(0)}$ is chosen:
\bd
E_{-1}[i-2/3] F_{-1}[j_1-2/3] F_{-1}[j_2-2/3]
 F_{-1}[j_3-2/3],
\ed
for $i \geq 0$,
\bd
F_{-1}[j_1-2/3] F_{-1}[j_2-2/3] F_{-1}[j_3-2/3]E_{-1}[i-2/3],
\ed
for $i<0$,
\bd
H_{-2}[i-4/3] F_{-1}[j_1-2/3]F_{-1}[j_2-2/3],
\ed
for $i \geq 0$,
\bd
F_{-1}[j_1-2/3]F_{-1}[j_2-2/3] H_{-2}[i-4/3],
\ed
for $j$,
\bd
F_{-1}[j-2/3].
\ed
This family of terms is free. The identities obtained by the
 elements in the expression of \ref{qSerre} are the following ones:

for $E_{-1}[i-2/3] F_{-1}[j_1-2/3] F_{-1}[j_2-2/3]
 F_{-1}[j_3-2/3]$, and $F_{-1}[j_1-2/3] F_{-1}[j_2-2/3]
 F_{-1}[j_3-2/3]E_{-1}[i-2/3]$,

\bd
 (V_{-} [-i-1] V_{+}[-j_1-1]V_{+}[-j_2-1]V_{+}[-j_3-1] -
\ed
\bd
-[3] V_{+}[-j_1-1] V_{-}[-i-1]V_{+}[-j_2-1]V_{+}[-j_3-1] +
\ed
\bd
+ [3] V_{+}[-j_1-1]V_{+}[-j_2-1]V_{-}[-i-1]V_{+}[-j_3-1] -
\ed
\bd
-V_{+}[-j_1-1]V_{+}[-j_2-1]V_{+}[-j_3-1]V_{-}[-i-1])=0,
\ed
because of the $q$-Serre relation between $V_\pm$.

For the other expressions, the terms are then put in the form
 of contour integrals. $z_2$ from $1$ to $1$ rotating around
 $0$ in the trigonometric way, $z_1$  too, around $z_2$,
  $z$ around $z_1$ and $z_2$, and endly $z'$ around $z_1$, $z_2$ and $z$.

\bd \label{continu}
 - \int V_-(z')V_+(z)V_+(z_1)V_+(z_2)[1/(z-z')][z^{i-1}z_1^{j_1}z_2^{j_2}]
 dz dz' dz_1 dz_2-
\ed
\bd
-\int V_-(z')V_+(z_1)V_+(z)V_+(z_2)[1/(z-z')][z^{i-1}z_1^{j_1}z_2^{j_2}]
 dz dz' dz_1 dz_2-
\ed
\bd
-\int V_-(z')V_+(z_1)V_+(z_2)V_+(z)[1/(z-z')][z^{i-1}z_1^{j_1}z_2^{j_2}]
 dz dz' dz_1 dz_2+
\ed
\bd
+[3] \int V_+(z)V_-(z')V_+(z_1)V_+(z_2)[1/(z-z')][z^{i-1}z_1^{j_1}z_2^{j_2}]
 dz dz' dz_1 dz_2+
\ed
\bd
+[3] \int V_+(z_1)V_-(z')V_+(z)V_+(z_2)[1/(z-z')][z^{i-1}z_1^{j_1}z_2^{j_2}]
 dz dz' dz_1 dz_2+
\ed
\bd
+[3] \int V_+(z_1)V_-(z')V_+(z_2)V_+(z)[1/(z-z')][z^{i-1}z_1^{j_1}z_2^{j_2}]
 dz dz' dz_1 dz_2-
\ed
\bd
-[3] \int V_+(z_1)V_+(z)V_-(z')V_+(z_2)[1/(z-z')][z^{i-1}z_1^{j_1}z_2^{j_2}]
 dz dz' dz_1 dz_2-
\ed
\bd
-[3] \int V_+(z)V_+(z_1)V_-(z')V_+(z_2)[1/(z-z')][z^{i-1}z_1^{j_1}z_2^{j_2}]
 dz dz' dz_1 dz_2-
\ed
\bd
-[3] \int V_+(z_1)V_+(z_2)V_-(z')V_+(z)[1/(z-z')][z^{i-1}z_1^{j_1}z_2^{j_2}]
 dz dz' dz_1 dz_2+
\ed
\bd
+\int V_+(z_1)V_+(z_2)V_+(z)V_-(z')[1/(z-z')][z^{i-1}z_1^{j_1}z_2^{j_2}]
 dz dz' dz_1 dz_2+
\ed
\bd
+\int V_+(z_1)V_+(z)V_+(z_2)V_-(z')[1/(z-z')][z^{i-1}z_1^{j_1}z_2^{j_2}]
 dz dz' dz_1 dz_2+
\ed
\bd
+\int V_+(z)V_+(z_1)V_+(z_2)V_-(z')[1/(z-z')][z^{i-1}z_1^{j_1}z_2^{j_2}]
 dz dz' dz_1 dz_2+
\ed
\bd
+(z_1 \leftrightarrow z_2)=0.
\ed
\bd
\int V_-(z')V_+(\xi) V_+(z)V_+(z_1)[1/(z-z')][1/(\xi-z)] z_1^{j}
 dzdz'd\xi dz_1 +
\ed
\bd
+\int V_-(z')V_+(z_1) V_+(\xi )V_+(z)[1/(z-z')][1/(\xi-z)] z_1^{j}
 dzdz'd\xi dz_1 +
\ed
\bd
+\int V_-(z')V_+(\xi) V_+(z_1)V_+(z)[1/(z-z')][1/(\xi-z)] z_1^{j}
 dzdz'd\xi dz_1 -
\ed
\bd
-[3](\int V_+(\xi) V_-(z')V_+(z)V_+(z_1)[1/(z-z')][1/(\xi-z)] z_1^{j}
 dzdz'd\xi dz_1 +
\ed
\bd
+\int V_+(z_1) V_-(z')V_+(\xi)V_+(z)[1/(z-z')][1/(\xi-z)] z_1^{j}
 dzdz'd\xi dz_1 +
\ed
\bd
+\int V_+(\xi) V_-(z')V_+(z_1)V_+(z)[1/(z-z')][1/(\xi-z)] z_1^{j}
 dzdz'd\xi dz_1) +
\ed
\bd
+[3](\int V_+(\xi) V_+(z)V_-(z')V_+(z_1)[1/(z-z')][1/(\xi-z)] z_1^{j}
 dzdz'd\xi dz_1 +
\ed
\bd
+\int V_+(z_1) V_+(\xi) V_-(z')V_+(z)[1/(z-z')][1/(\xi-z)] z_1^{j}
 dzdz'd\xi dz_1 +
\ed
\bd
+\int V_+(\xi) V_+(z_1)V_-(z')V_+(z)[1/(z-z')][1/(\xi-z)] z_1^{j}
 dzdz'd\xi dz_1 ) -
\ed
\bd
-(\int V_+(z_1) V_+(\xi)V_+(z)V_-(z')[1/(z-z')][1/(\xi-z)] z_1^{j}
 dzdz'd\xi dz_1 +
\ed
\bd
+\int V_+(\xi) V_+(z_1) V_+(z)V_-(z')[1/(z-z')][1/(\xi-z)] z_1^{j}
 dzdz'd\xi dz_1 +
\ed
\bd
+\int V_+(\xi)V_+(z) V_+(z_1)V_-(z')[1/(z-z')][1/(\xi-z)] z_1^{j}
 dzdz'd\xi dz_1 ) + [
\ed
\bd
- (\int V_-(z') V_+(z)V_+(\xi)V_+(z_1)[1/(z-z')][1/(\xi-z)] z_1^{j}
 dzdz'd\xi dz_1 +
\ed
\bd
+\int V_-(z') V_+(z)V_+(z_1)V_+(\xi)[1/(z-z')][1/(\xi-z)] z_1^{j}
 dzdz'd\xi dz_1 +
\ed
\bd
+\int V_-(z') V_+(z_1) V_+(z) V_+(\xi)[1/(z-z')][1/(\xi-z)] z_1^{j}
 dzdz'd\xi dz_1) +
\ed
\bd
+[3] \int V_+(z) V_-(z')V_+(\xi)V_+(z_1)[1/(z-z')][1/(\xi-z)] z_1^{j}
 dzdz'd\xi dz_1 +
\ed
\bd
+[3]\int  V_+(z)V_-(z') V_+(z_1)V_+(\xi)[1/(z-z')][1/(\xi-z)] z_1^{j}
 dzdz'd\xi dz_1 +
\ed
\bd
+[3] \int  V_+(z_1)V_-(z') V_+(z)V_+(\xi)[1/(z-z')][1/(\xi-z)] z_1^{j}
 dzdz'd\xi dz_1 -
\ed
\bd
-[3]\int  V_+(z_1)V_+(z) V_-(z')V_+(\xi)[1/(z-z')][1/(\xi-z)] z_1^{j}
 dzdz'd\xi dz_1 -
\ed
\bd
-[3] \int  V_+(z)V_+(\xi)V_-(z') V_+(z_1)[1/(z-z')][1/(\xi-z)] z_1^{j}
 dzdz'd\xi dz_1 +
\ed
\bd
-[3] \int  V_+(z)V_+(z_1)V_-(z') V_+(\xi)[1/(z-z')][1/(\xi-z)] z_1^{j}
 dzdz'd\xi dz_1 -
\ed
\bd
-(-\int  V_+(z)V_+(\xi)V_+(z_1)V_-(z') [1/(z-z')][1/(\xi-z)] z_1^{j}
 dzdz'd\xi dz_1 -
\ed
\bd
-\int  V_+(z_1)V_+(z)V_+(\xi)V_-(z') [1/(z-z')][1/(\xi-z)] z_1^{j}
 dzdz'd\xi dz_1 -
\ed
\bd
-\int  V_+(z)V_+(z_1)V_+(\xi)V_-(z') [1/(z-z')][1/(\xi-z)] z_1^{j}
 dzdz'd\xi dz_1)]=0.
\ed
 If the first expression vanishes for $V_+(z)$ and $V_-(z')$
 in the case of the zero modes, the same manipulation implies the
 vanishing for all the $\al_+(z) V_+(z)$ and
 $\al_-(z') V_-(z')$, using for $\al_+$ and $\al_-$ powers of $z,
 z'$. By linear combination, it implies the general equality.
 We will therefore prove the identity for zero modes. 

To show the vanishing, the contours are deplaced toward the unity circle
 and each of the integrals are decomposed with the angular sectors.

\bd
 \int_S V_-(z')V_+(z)V_+(z_1)V_+(z_2)[1/(z-z')] dz dz' dz_1 dz_2,
\ed
with $S$ a certain angular sector.

{\lem .
For the contour integrals, the rule of calculus is the following one, for
 $0< arg(z)< arg(z')< 2 \pi$ and $|z| > |z'|$:

\bd
V_+ (z) V_-(z')= ( z - z')^{\al}: V_+(z) V_-(z'):,
\ed
it gives:
\bd
V_+(z) V_-(z') = e^{i \al \pi} V_-(z')V_+(z),
\ed
with:
\bd
q^{-1}=e^{i \al \pi}, \al = \be^2.
\ed}

The vanishing of the coefficients must be showed with each of
 the angular sectors, using the standard integrals by the rules of calculus.
For this aim, it is made use of a lattice. Because
 of the independance of the angular sectors for a lattice
 (lemma \ref{independance}), the vanishing of the coefficients
for the angular sectors of the lattice implies that the coefficients
 vanish for the ones of the contour integrals, and so proves
 the identity.

\subsection{The lattice}

The lattice is given by $\ZZ$ and the variables are $x^+_i, x^-_i$, with:
\bd
x^e_i x^{e'}_{i'} = q^{ee'} x^{e'}_{i'}x^e_i,
\ed
for $i<i'$, $i,i' \in \{ 1, 2, ..., N \}$ and $e,e'=+,-$;
it allows to give a definition of the dicret analogue of the charge operators, 
\bd
V^d_\pm = \sum_{i=1}^N x_i^\pm.
\ed

 For a permutation of $\{ z, w_1, w_2, w_3 \}$, $\sa$, and for
 a fixed angular sector, the integration over it is:

\bd
\int_S V_{sig(\sa(z))}(\sa(z)) V_{sig(\sa(w_1))}(\sa(w_1)) 
 V_{sig(\sa(w_2))}(\sa(w_2) V_{sig(\sa(w_3))}(\sa(w_3))
\ed
\bd
= c(\sa)  \int_{\sa^{-1}(S)} V_-(z) V_+(w_1) V_+(w_2) V_+(w_3),
\ed
with $sig$, the sign corresponding with the permutated term, 
and with $c(\sa)$, a coefficient which depends on the permutation.

Also for the permutation of $\{ i, j_1, j_2, j_3 \}$, $\sa$, the sum
 over the corresponding lattice sector is:

\bd
\sum_{i, j_k \in S^r}
 x^{sig(\sa(i)}_{\sa(i)} x^{sig(\sa(j_1)}_{\sa(j_1)}
 x^{sig(\sa(j_2)}_{\sa(j_2)} x^{sig(\sa(j_3)}_{\sa(j_3)}=
\ed 
\bd
=c(\sa) \sum_{i, j_k \in 
\sa^{-1}(S^r)} x_i^- x_{j_1}^+   x_{j_2}^+ x_{j_3}^+,
\ed
the rules of permutation are the same.

The vanishing of the coefficients of the lattice sums is proved
 using an identity for the lattice (\ref{reseau})
 (lemma \ref{idres}), after having showed the independance
 of the angular sectors over the lattice (lemma \ref{independance}).
 The vanishing of the coefficients of the standard integrals over
 the sectors is showed and the identity (\ref{continu}).

 the sectors is showed and the identity (\ref{continu}).

{ \lem. \label{independance}
The angular sectors over the lattice are independant.}

{\lem . The following identity for the lattice is satisfied: \label{idres}
\bd \label{reseau}
 \sum_{j,j',j_1,j_2} - x_{j'}^- x_j^+ x_{j_1}^+ x_{j_2}^+ e(j,j')-
  x_{j'}^- x_{j_1}^+ x_j^+  x_{j_2}^+ e(j,j')-
\ed
\bd
-
  x_{j'}^- x_{j_1}^+ x_{j_2}^+ x_j^+ e(j,j')+
\ed
\bd
+[3](\sum_{j,j',j_1,j_2} x_j^+ x_{j'}^- x_{j_1}^+ x_{j_2}^+ e(j,j')+
  x_{j_1}^+ x_{j'}^- x_{j}^+ x_{j_2}^+  e(j,j')+
\ed
\bd
+
  x_{j_1}^+ x_{j'}^- x_{j_2}^+ x_{j}^+  e(j,j'))+
\ed
\bd
+[3](\sum_{j,j',j_1,j_2} - x_{j_1}^+ x_{j}^+  x_{j'}^- x_{j_2}^+  e(j,j')-
  x_{j}^+ x_{j_1}^+  x_{j'}^- x_{j_2}^+  e(j,j')-
\ed
\bd
-
  x_{j_1}^+ x_{j_2}^+  x_{j'}^- x_{j}^+  e(j,j'))+
\ed
\bd
+\sum_{j,j',j_1,j_2} x_{j_1}^+ x_{j_2}^+  x_{j}^+ x_{j'}^-  e(j,j')+
 x_{j_1}^+ x_{j}^+  x_{j_2}^+ x_{j'}^-  e(j,j')+
\ed
\bd
+
 x_{j}^+ x_{j_1}^+  x_{j_2}^+ x_{j'}^-  e(j,j')+
\ed
\bd
+(j_1 \leftrightarrow j_2) =0.
\ed}

{\it Proof}:

 $e$ is decomposed in elementary antisymmetric functions;
 so we can have $e(j,j')=1$ and the other terms are zero.

Some derivations of the ring $\CC[x^{\pm}_i]$ are used:

{\lem . The following  formulas define derivations of the ring 
$\CC[x^{\pm}_i]$:
\bd          
\da_i^+, \da_i^-,
\ed          
\bd          
\da_i^+ (x_j^{\ep})= \da_{\ep +} \da_{ij} x_j^{\ep},
\ed          
\bd          
\da_i^-(x_j^{\ep})= \da_{\ep -} \da_{ij} x_j^{\ep}.
\ed}
                  
The derivation of the $q$-Serre relation $S^r=0$ for the lattice,
 which is proved by use of the coproduct of $U_q \widehat {\SL}_2$, is then:
    
\bd          
S^r= V_- V_+^3 -[3] V_+V_-V_+^2 +[3]V_+^2V_-V_+-V_+^3V_-=0,
\ed          
\bd          
\da_j^+ \da_{j'}^- (S) = \da_j^+ ( x_{j'}^- V_+^3 -[3] V_+ x_{j'}^- V_+^2
 +[3] V_+^2 x_{j'}^- V_+ - V_+^3 x_{j'}^-)=
\ed          
\bd          
= \sum_{j_1,j_2} - x_{j'}^- x_j^+ x_{j_1}^+ x_{j_2}^+ -
  x_{j'}^- x_{j_1}^+ x_j^+  x_{j_2}^+ -
  x_{j'}^- x_{j_1}^+ x_{j_2}^+ x_j^+ +
\ed          
\bd          
+ [3](\sum_{j_1,j_2} x_j^+ x_{j'}^- x_{j_1}^+ x_{j_2}^+ +
  x_{j_1}^+ x_{j'}^- x_{j}^+ x_{j_2}^+ +
 x_{j_1}^+ x_{j'}^- x_{j_2}^+ x_{j}^+  +
\ed          
\bd          
+ [3](\sum_{j_1,j_2} - x_{j_1}^+ x_{j}^+  x_{j'}^- x_{j_2}^+  -
  x_{j}^+ x_{j_1}^+  x_{j'}^- x_{j_2}^+  -
  x_{j_1}^+ x_{j_2}^+  x_{j'}^- x_{j}^+  +
\ed          
\bd          
+ \sum_{j_1,j_2} x_{j_1}^+ x_{j_2}^+  x_{j}^+ x_{j'}^- +
 x_{j_1}^+ x_{j}^+  x_{j_2}^+ x_{j'}^-  +
 x_{j}^+ x_{j_1}^+  x_{j_2}^+ x_{j'}^- +
\ed          
\bd          
+ (j_1 \leftrightarrow j_2) =0.
\ed       
The identity (\ref{reseau}) is obtained.

$\Box$

A same calculation holds in the case $F_{-1}$, the derivations
 $\da_i^\pm$ are applied three times.

\section{Calculations of cohomology}

\subsection{The quantum resolution BGG}

The quantum resolution BGG ($q$-BGG) is constructed with Verma modules 
\cite{FF}:

\bd
B^q_j(\tilde {\mf g}) = \oplus_{l(s)=j}M^q_{\rho-s(\rho)},
\ed
 
with $M^q_{\rho-s(\rho)}$ a module of Verma.

{\defi .
\bd
\tilde {\cal F}_0^{\beta}= \tilde \pi_0^{\beta},
\ed

if $n \neq 0$,
\bd
\tilde {\cal F}_n^{\beta}= \tilde \pi_n^{\beta} \oplus \tilde \pi_{-n}^{\beta}.
\ed}

{\defi . 
A degree $d_{VOA}$ is defined for $\tilde {\cal F}^{\beta}_n$,
 $v_n^{\beta}$ being the vector of highest weight:

\bd
d_{VOA} v_n^{\beta}= n^2/3,
\ed
\bd
d_{VOA}X[i]=i.
\ed}

$d$ respects the degrees.

{\nota . For all $n$, define $\tilde H_{k}(n)^{\be}$ as
the cohomology $H_{k}(\tilde {\cal F}^{\beta}(n))$.}

\subsection{The degrees of the $\tilde \pi_n$}

{\lem . $\bar V_\pm$ are morphisms of $\tilde \pi^{\beta}_n$
  towards $\tilde \pi^{\beta}_{n+1}$,
 if $\tilde \pi^{\beta}_n$ has for degree VOA $d_{VOA}-n^2/3$ and
 $\tilde \pi^{\beta}_{n+1}$, $d_{VOA}-(n+1)^2/3$, then they are
 of degree $1/3$.}

At level of $q$-BGG, the cohomology of the resolution is ${\La}^* {\mf a}^*$.

The degree of the resolution gives
$\tilde \pi_0^{\be}$ of degree $0$, $\tilde \pi_1^{\be}$ is of degree $1$,
 $\tilde \pi_{-1}^{\be}$ of degree $-2/3$. The differential is of degree $0$.

\subsection{The character of the resolution}

Let us fix $n \in [1/3] \ZZ$.  The Euler characteristic of the resolution
 is for the degree $n$ component of the VOA:

\bd
{\chi}^{\beta}(n)(\tilde {\cal F}^{\beta}_*, d) = \sum_i dim(\tilde
{\cal F}^{\beta}_{2i}(n)) - \sum_i dim(\tilde {\cal F}^{\beta}_{2i+1}(n)) =
\ed
\bd
= \sum_i dim(\tilde {H}_{2i}(n)^{\be}) -
 \sum_i dim(\tilde {H}_{2i+1}(n)^{\be}).
\ed

The character is the same in classical as in quantum. The character
 of the quantum cohomology is the same as that of the quantum complex,
 as the character of the cohomology is the one of the complex,
 it is obtained that the character is the same in classical as in quantum.

The classical cohomology is zero in odd degree
if $3n$ is even, $n$ being $d_{VOA}$, and zero in even degree if $3n$ is odd.

In a point of specialisation, the cohomology increases; so that:

\bd
dim(\tilde H_{2i}(n)^{\be}) \leq dim(\tilde H_{2i}(n)),
\ed

\bd
dim(\tilde H_{2i+1}(n)^{\be}) \leq dim(\tilde H_{2i+1}(n)).
\ed

As:

\bd
dim(\tilde H_{2i+1}(n))=0,
\ed

\bd
dim(\tilde H_{2i+1}(n)^{\be})=0.
\ed

And:
\bd
dim(\tilde H_{2i}(n)^{\be}) \leq dim(\tilde H_{2i}(n)).
\ed

As the character is the same in quantum as in classical:

\bd
\sum_i dim(\tilde H_{2i}(n)^{\be})= \sum_{i} dim(\tilde H_{2i}(n)),
\ed

it is obtained:

\bd
dim(\tilde H_{2i}(n)^{\be}) = dim(\tilde H_{2i}(n)).
\ed

Spectral sequence \cite{G} must be considered, the vertical cohomology is
 $\La^* {\mf a}^*$, with a horizontal differential
 of degree $1$. The vertical differential of the spectral sequence
 is of degree $(0,1)$, the horizontal one is of degree $(1,0)$.

The cohomology of the total complex is $E_2^{p,q}$.

For $p=0,1$, $E_2^{p,q}= \La^q {\mf a}^*$, for $p \neq 0,1$, $E_2^{p,q}=0$.

The cohomology of the total complex is $\sum_{p+q=n}E_2^{p,q}$.
The $H^1$ of the total complex is
 $E_2^{1,0} \oplus E_2^{0,1}$, it gives, $\CC \oplus {\mf a}^*$.

\subsection{The quantum integrals of motion}

{\th .  \label{intmot}
The integrals of motion $\oint d_{2i+1}$ admit quantum deformation, which
 belong to a completed universal envelopping algebra and which
 commute with each other and have the same degrees as the integrals
 of motion.} 

\section{Acknowledgements}

I greatly thank B.Enriquez for his contribution to this article,
 his advices and encouragements; and C.Grunspan for helpful discussions.
I must mention also Y.Kosmann-Schwarzbach for guidance in Lie algebra theory.

\end{document}